\documentclass[a4paper,11pt]{article}
\usepackage{amsmath,amssymb,bm,epsfig,color,cite,braket,mathrsfs,slashed,multirow,graphicx,latexsym,cancel,comment}

\renewcommand{\thefootnote}{\fnsymbol{footnote}}

\usepackage{hyperref,xcolor}
\hypersetup{
colorlinks=true,
linktocpage=true,
linkcolor=blue,
filecolor=blue,      
urlcolor=blue,
citecolor=blue,
   }
    
\begin{document}
\title{
\begin{flushright}
 \begin{minipage}{0.3\linewidth}
 \normalsize
KCL-PH-TH-2024-51\\ UT-Komaba/23-11
\vspace{+.3cm}
 \end{minipage}
\end{flushright}
{\Large \bf Formation of defects associated with both \\ spontaneous and explicit symmetry breaking\\*[20pt]}}

\author{
Kohei Fujikura$^{1}$,\footnote{
\href{mailto:}{kfujikura@g.ecc.u-tokyo.ac.jp}}\ \
Mairi Sakellariadou$^{2}$,\footnote{
\href{mailto:}{mairi.sakellariadou@kcl.ac.uk}}\ \
Michiru Uwabo-Niibo$^{3,4}$,\footnote{
\href{mailto:g2370609@edu.cc.ocha.ac.jp}{g2370609@edu.cc.ocha.ac.jp}}
\ and
Masahide Yamaguchi$^{4,5}$\footnote{
\href{mailto:}{gucci@ibs.re.kr}}
\\*[20pt]
$^1${\it \small
Graduate School of Arts and Sciences, University of Tokyo,}
\\{\it\small Komaba, Meguro-ku, Tokyo 153-8902, Japan} \\
$^{2}${\it \small
Theoretical Particle Physics and Cosmology Group, Physics Department,
King’s College London,}
\\{\it\small University of London, Strand, London WC2R 2LS, United Kingdom} 
\\
$^3${\it \small
Department of Physics, Graduate School of Humanities and Sciences,}\\{\it\small  Ochanomizu University, Tokyo 112-8610, Japan} 
\\
$^4${\it\small
Cosmology, Gravity, and Astroparticle Physics Group, Center for
Theoretical Physics of the Universe, }\\{\it\small  Institute for Basic Science (IBS), Daejeon,
34126, Korea}
\\
$^{5}${\it \small
Department of Physics, Tokyo Institute of Technology, Tokyo, 152-8551, Japan}
\\{\it\small } 
\\*[50pt]
}

\date{
\centerline{\small \bf Abstract}
\begin{minipage}{0.9\linewidth}
\medskip \medskip \small 
We discuss formation of cosmic strings associated with a spontaneously broken {\it approximate} $U(1)$ symmetry by performing classical field-theoretical simulations.
An original $U(1)$ symmetry is explicitly broken down to its subgroup $Z_N$ even before spontaneous breaking takes place.
 We estimate the ratio of explicit breaking to that of spontaneous breaking for which topological defects for $N=1$ and $N=2$ are formed.
For $N=1$, a cosmic string attached to a single domain wall can be formed when the amount of the explicit breaking is three orders of magnitude smaller than that of the spontaneous breaking.
For $N=2$, no matter how large the explicit breaking is, domain walls are inevitably formed as long as the temperature of the Universe is high enough to restore $Z_2$ symmetry.
In that case, cosmic strings are also inevitably formed as long as the amount of the explicit breaking is smaller than that of the spontaneous breaking.
\end{minipage}
}

\maketitle{}
\thispagestyle{empty}
\addtocounter{page}{-1}
\clearpage
\noindent
\hrule
\tableofcontents
\noindent
\hrulefill

\renewcommand{\thefootnote}{\arabic{footnote}}
\setcounter{footnote}{0}

\section{Introduction}
\label{sec:introduction}

Topological defects can be formed when topology of the ground state is non-trivial.
In particular, when a thermal phase transition associated with $U(1)$ symmetry breaking takes place, the topology of the ground state is not simply-connected, resulting in the formation of one-dimensional localized configuration called cosmic strings by the Kibble-Zurek mechanism~\cite{Kibble:1976sj,Zurek:1985qw}. Cosmic strings are expected to be generically formed in the context of Grand Unified Theories~\cite{Jeannerot:2003qv}.
Since a continuum deformation does not change topology of the ground state, cosmic strings are stable once they are formed.
Hence cosmic strings are generically long-lived and lead to many interesting phenomena in late-time cosmology such as the production of gravitational wave and dark matter (see e.g, Ref.~\cite{Vilenkin:2000jqa} for the review of the conventional cosmic string dynamics and its phenomena).

In some theories beyond the standard model (SM) of particle physics, an additional approximate global $U(1)$ symmetry, which is spontaneously broken at a certain energy scale, is introduced.
For example, when there exists chiral global $U(1)$ symmetry, which is anomalous under the SM $SU(3)_C$ gauge symmetry, is spontaneous broken, a pseudo-Nambu-Goldstone boson (pNGB) called axion appears.
An axion obtains its potential via the non-perturbative QCD instanton effect and dynamically solves the strong CP problem in the QCD~
\cite{Peccei:1977hh,Weinberg:1977ma,Wilczek:1977pj,Vafa:1984xg}.
Another interesting example is a majoron model.
In this latter model, right-handed neutrinos are introduced and an approximate global $U(1)_L$ symmetry associated with the lepton number is assumed to be spontaneously broken by a condensation of a gauge singlet scalar carrying lepton number~\cite{Chikashige:1980ui,Schechter:1981cv}.
After the spontaneous breaking of $U(1)_L$, the pNGB called majoron appears and can be a suitable candidate of a cold dark matter with an adequate mass obtained through a certain amount of the explicit breaking of $U(1)_L$~\cite{Rothstein:1992rh,Berezinsky:1993fm,Lattanzi:2007ux,Bazzocchi:2008fh,Frigerio:2011in,Queiroz:2014yna,Reig:2019sok}.

In both cases, an additional global $U(1)$ symmetry is regarded as an approximate symmetry for phenomenological reasons.
Even for theoretical perspectives, it is also conceivable that there are no exact continuous global symmetries in the UV-complete quantum gravity such as string theory~\cite{Giddings:1989bq,Gibbons:1995vg,Rey:1989xj,Gutperle:2002km,Bergshoeff:2004fq,Hebecker:2016dsw,Harlow:2018jwu,Harlow:2018tng}.
Indeed, non-perturbative quantum gravity effects such as the creation of Euclidean wormhole explicitly would break any global $U(1)$ symmetries~\cite{Giddings:1987cg,Lee:1988ge,Giddings:1988cx,Coleman:1988cy,Abbott:1989jw,Kallosh:1995hi,Yonekura:2020ino}. (For review on the wormholes, see e.g. Ref.~\cite{Hebecker:2018ofv}.)
Therefore it is natural to assume the presence of an explicit breaking of global $U(1)$ symmetry in a realistic setup.
These explicit breaking effects could change the fate of topological defects in the early Universe.

The evolution of cosmic strings associated with the {\it subsequent} explicit breaking of an original spontaneously broken $U(1)$ symmetry has been investigated so far in the context of the axion string, where explicit breaking is dominated by the QCD instanton effect~\cite{Sikivie:1982qv,Vilenkin:1982ks,Vilenkin:1984ib}.
It is known that cosmological consequences heavily depend on the type of explicit breaking of the $U(1)$ symmetry.
Suppose that there remains a discrete subgroup $Z_N$ ($N \in \mathbb{N}$) of the original $U(1)$ symmetry.
For $N=1$, it is argued in Refs.~\cite{Vilenkin:1982ks,Vilenkin:1984ib}, that there are transitions from a string network to a string-wall network, in which domain walls are connected by strings.
The string-wall system soon collapses by the domain wall tension~\cite{Hiramatsu:2010yn,Hiramatsu:2012gg}, once the Hubble parameter becomes comparable to the typical energy scale of the domain wall.
For $N>1$, a string-wall network is realized in a similar way as for the $N=1$ case.
However, for $N>1$ domain walls are stable, in contrary to the $N=1$ case because there exists $N$ disconnected degenerate ground states~\cite{Sikivie:1982qv}.
Stable domain walls soon dominate the energy density of the Universe which is in conflict with cosmological observations~\cite{Zeldovich:1974uw}.
Field configurations and the dynamics of such composite solitons are also recently studied in Ref.~\cite{Eto:2023gfn}.

One usually implicitly assumes that cosmic strings are formed via the ordinary Kibble-Zurek mechanism even in the presence of explicit breaking.
This assumption may be trivially justified in the case of invisible axion models~\cite{Kim:1979if,Shifman:1979if,Zhitnitsky:1980tq,Dine:1981rt}
because QCD instanton effects are negligible at the epoch of string formation due to the large hierarchy between the explicit and spontaneous symmetry breaking scales. We will quantitatively discuss this point in Sec.~\ref{sec:numerical simulation}.
However, it is  non-trivial whether cosmic strings are formed or not in a generic setup such as the majoron model, since the explicit breaking effect is not necessarily many magnitude smaller than that of spontaneous breaking.\footnote{Note that there may be an explicit breaking of a global $U(1)$ symmetry even in invisible axion models from quantum gravity in addition to the ordinary QCD instanton effect. However, such a contribution generically reintroduces the strong CP problem.}
In the following, we will investigate cosmic string formation associated with spontaneously broken {\it approximate} global $U(1)$ symmetry.
We begin with the simplest Goldstone model~\cite{Goldstone:343400}, where global $U(1)$ symmetry is spontaneously broken by a condensation of a complex scalar field.
Then, we introduce relevant operators that break the original $U(1)$ down to its discrete subgroup $Z_N$, which is similar to the case of invisible axion models and majoron models.
In both cases, it is known that there is a composite soliton configuration where a cosmic string is attached to $N$ domain wall(s) when the effect of explicit breaking is subdominant compared to that of the spontaneous breaking.
However, as we will argue, the production mechanism of such defects is qualitatively different for $N=1$ and $N=2$.

For $N=1$, there is no phase transition associated with breaking of $U(1)$ since there is no exact symmetry.
In fact, the unique 
ground state is realized during the cosmological evolution so that there is no formation of cosmic strings within a mean field approximation. (See Fig.~\ref{fig:potentail-n1} for the explanation of no cosmic string formation with a mean field approximation.)
However, there may be  thermal fluctuations in addition to the primordial fluctuation created  in the early Universe.
Hence, if the effect of fluctuations overcomes that of the explicit symmetry breaking, the ordinary Kibble-Zurek mechanism may operate leading to cosmic strings formation.
In the opposite case where the effect of explicit breaking is large enough to make field configurations trivial, one does not expect cosmic strings formation.
In order to quantify a size of the explicit symmetry breaking leading to formation of cosmic string, we perform the classical field-theoretic simulation in the radiation-dominated expanding Universe.

For $N=2$, there is a phase transition associated with spontaneously broken $Z_2$ symmetry in a strict sense.
Therefore, no matter how large the explicit breaking is, domain walls are inevitably formed as long as the temperature of the Universe is high enough to restore the $Z_2$ symmetry.
After the spontaneous breaking of $Z_2$ symmetry, two disconnected ground states are randomly realized in the Universe, unless there is no bias to choose one of them.
One may expect that the presence of randomness of two disconnected ground states and smoothness of the field configuration may lead to formation of composite solitons, where cosmic strings are attached to two domain walls.
Using classical lattice field simulations, we will investigate whether a string-wall network is actually formed and whether it is stable as long as the size of the explicit symmetry breaking is smaller than that of the spontaneous symmetry breaking, contrary to the $N =1$ case.

This paper is organized as follows.
We present our setup and defects formation under explicit symmetry breaking of a global $U(1)$ symmetry in Sec.~\ref{sec:setup}.
Some speculative discussions on the formation of defects are presented including effects of thermal inhomogeneous fluctuations.
In Sec.~\ref{sec:numerical simulation}, we describe our setup of the classical field simulation.
Results of simulations are given in Sec.~\ref{sec:result}.
Section \ref{sec:discussion} is devoted to conclusions and discussion.


\section{The model}\label{sec:setup}

Let us present our model and analyze cosmological evolution of the ground state with and without an explicit symmetry breaking by taking thermal effects into account.

We consider a spatially flat isotropic Friedmann‐Lema\^{i}tre‐Robertson-Walker (FLRW) spacetime
\begin{align}
    \mathrm{d}s^{2}=\mathrm{d}t^{2}-a(t)^{2}(\mathrm{d}x^{2}+\mathrm{d}y^{2}+\mathrm{d}z^{2}), \label{eq:metric}
\end{align}
with $a(t)$ the scale factor.
We consider the simplest Goldstone model~\cite{Goldstone:343400} with an additional explicit breaking interaction, whose Lagrangian density is defined as
\begin{align}
&\mathcal{L}=\partial_{\mu}\Phi^{\dagger}\partial^{\mu}\Phi-\mathcal{V}(\Phi),~\mathcal{V}(\Phi) = \mathcal{V}_{U(1)}(|\Phi|)+\mathcal{V}_{\cancel{U(1)}}\,(\Phi),\nonumber\\
\mbox{where\ \ } &\mathcal{V}_{U(1)}(|\Phi|) = \dfrac{\lambda}{2}\left(|\Phi|^2-\sigma^2\right)^{2},\\
&\mathcal{V}_{\cancel{U(1)}}(\Phi) =\frac{1}{N}\lambda b_{N}\sigma^{4-N}(\Phi^{N}+{\rm c.c.}),~(N\in \mathbb{N})\nonumber
\end{align}
with $\Phi$ a complex scalar field.
Note that although $\Phi$ can be coupled to other matter fields including SM fields, in our analysis we ignore them for simplicity.
The $\mathcal{V}_{U(1)}$ respects the global $U(1)$ symmetry whose transformation property is defined by $\Phi \to e^{i\theta}\Phi$. The
$\mathcal{V}_{\cancel{U(1)}}$ breaks it down to its discrete group $Z_2^{\rm CP}\times Z_N$, where $Z_2^{\rm CP}$ is the CP symmetry whose transformation property is defined by $\Phi\to \Phi^*$~\cite{Eto:2023gfn}.
The dimensionless parameter $b_N$ quantifies the ratio of explicit breaking to that of the spontaneous breaking.
In the following, we assume that $b_N$ is a constant satisfying $0\leq b_{N}<1$ and does not depend on the cosmic time. (In the invisible axion model, this parameter depends on the cosmic time since the explicit breaking arises from the QCD instanton depending on the temperature. We will shortly discuss this point in Sec.~\ref{sec:explicit breaking effects}.)
One can make $b_N$ positive by a suitable redefinition of $\Phi$.
Although one may need to sum up all possible explicit breaking terms $N=1,2,\cdots$ in the completely generic discussion, we will focus on two cases, $N=1$ and $N=2$.
Inclusion of higher order terms and/or the mixed case where two explicit breaking terms with $N=1$ and $N=2$ coexist, is more complicated and left for a future study.

We now include a finite-temperature effect by assuming that $\Phi$ is in thermal equilibrium.
An effect of the finite-temperature fluctuation can be captured using the ordinary imaginary time formulation (See e.g. Refs.~\cite{Dolan:1973qd,Quiros:1999jp,Kapusta:2006pm} for the review of the finite-temperature effect and its applications).
In the high-temperature limit ($T\gg \Phi$) at one-loop order, a total effective potential of $\Phi$ can be expressed as
\begin{align}
    &\mathcal{V}(\Phi,T)=\mathcal{V}_{\rm SSB}(\Phi,T)+\mathcal{V}_{\cancel{U(1)}}(\Phi),\nonumber\\
    \mbox{where\ \ }&\mathcal{V}_{\rm SSB}(\Phi,T)= \frac{1}{2}\lambda\left(\Phi^{\dagger}\Phi-\eta^{2}(T)\right)^{2}~~\mbox{with\ \ }\eta^{2}(T)=\sigma^{2}-\frac{1}{3}T^{2};\label{eq:potential}
\end{align}
we omit contributions that do not depend on $\Phi$.

For a vanishing explicit symmetry breaking with $b_N =0$, the $U(1)$ symmetry is an exact symmetry.
In this case, a second-order phase transition takes place at the critical temperature $T=T_C$ defined by $T_C\equiv \sqrt{3}\sigma$.
The $U(1)$ symmetry is restored for high temperature $T>T_C$, characterized by $\langle \Phi\rangle =0$.
This symmetry is spontaneously broken $T<T_C$ by a condensation of $\langle \Phi\rangle \neq 0$.
In this case, the ground state can be parametrized by the phase of $\Phi$, as $\Phi=|\Phi| e^{i\theta}$, whose topology is identical to that of $\mathbb{S}^1$.
This non-trivial structure of the ground state admits the formation of cosmic strings when $\theta$ is randomly distributed at the epoch of the phase transition.
We will discuss the evolution of $\Phi$ based on the above effective potential including non-vanishing $b_N$ and then explain how this scheme changes for $N=1$ and $N=2$.

\subsection{The $N=1$ case}\label{subsec:setup-n1}

Let us focus on the  $N=1$ case.
The original $U(1)$ symmetry is explicitly broken to the trivial group, $Z_1$.
$Z_{2}^{\rm CP}$ symmetry is not spontaneously broken and hence there is only the trivial unique ground state.

One can rewrite the total effective potential $\mathcal{V}$ using the parametrization $\Phi=(\phi_1+i\phi_2)/\sqrt{2}$ as
\begin{align}
    \mathcal{V}(\phi_1,\phi_2,T)=\frac{1}{2}\lambda\left(\frac{1}{2}(\phi_{1}^{2}+\phi_{2}^{2})-\eta^{2}(T)\right)^{2}+ \sqrt{2}\lambda b_{1}\sigma^{3}\phi_{1}.
\end{align}
Clearly, $\partial \mathcal{V}/\partial \phi_1 =\partial\mathcal{V}/\partial \phi_2 =0$ imposes $\phi_2=0$.
Therefore it is sufficient to discuss 
evolution of $\phi_1$ by projecting the field space onto the $\phi_2=0$ line.

\begin{figure}
    \centering
    \includegraphics[width=10cm]{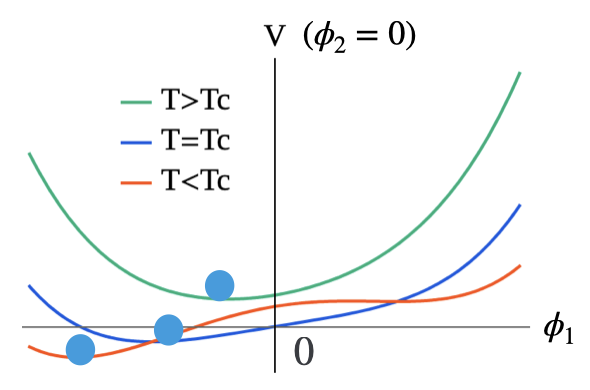}
    \caption{Total effective potentials with an explicit breaking $0<b_1<1$ projected onto the line $\phi_{2}=0$ are depicted for $T>T_C$ (green curve), $T=T_C$ (blue curve) and $T<T_C$ (red curve) in the case of $N=1$. Blue dots show the unique minimum of the total effective potential at each temperature. In the presence of the explicit breaking, $\phi_1$ develops a non-zero condensate at any temperature.}
    \label{fig:potentail-n1}
\end{figure}

Figure \ref{fig:potentail-n1} shows $\mathcal{V}(\phi_1,\phi_2=0,T)$ for temperatures $T>T_C,~T=T_C$ and $T<T_C$.
Due to the  linear term of $\phi_1$ in $\mathcal{V}$, $\partial \mathcal{V}/\partial \phi_1 = 0$ leads to $\langle \phi_1\rangle \neq 0$, no matter how high the temperature is.
For $T\gg T_C$, there is only one minimum of the potential, whose position is approximately given by 
\begin{align}
    \langle \phi_1\rangle = -\dfrac{3\sqrt{2}b_1\sigma^3}{T^2-T_C^2}+\mathcal{O}(b_1^3), ~(T \gg T_C).
\end{align}
We solve $\partial \mathcal{V}/\partial \phi_i = 0$ ($i=1,2$) by treating $b_1$ as a small perturbation.
The above expression does not hold for $T\simeq T_C$, since $b_1$ cannot be treated as a perturbation, and in this case 
we treat $ (T-T_C)/\sigma$ as a small perturbation.
The minimum of the potential minimum is 
\begin{align}
    \langle \phi_1\rangle = -\left(2\sqrt{2}b_1\right)^{1/3}\sigma+\mathcal{O}(T-T_C),~(T\simeq T_C).\label{eq:minimum at the critial temperature}
\end{align}
For $T \ll T_C$, we consider again  $b_1$ as a perturbation, and get the minimum of the potential at
\begin{align}
    \langle \phi_1\rangle = -\sqrt{\frac{2}{3}(T_C^2-T^2)}-\frac{3\sqrt{2}}{2}\frac{b_1\sigma^3}{T_C^2-T^2}+\mathcal{O}(b_1^2),~(T \ll T_C)\label{eq:<phi_1> for T<T_C}.
\end{align}
Note that $\phi_1$ evolves towards the unique minimum at any temperature, contrary to the case with vanishing $b_1$.
We numerically see that $\langle\phi_1\rangle$ is a smooth function with respect to $T$, which leads to a smooth crossover behavior of the free energy at least under a mean field analysis.
Therefore, there is no definite phase transition for $b_1\neq 0$, which may lead to a crossover behavior whose sharpness is controlled by $b_1$.

An approximate flat direction for $T\leq T_C$ emerges as long as we treat the explicit symmetry breaking term as the perturbation
\footnote{Precisely speaking, the condition to have three different solutions for $\partial \mathcal{V} /\partial\phi_{1}(\phi_{2}=0)=0 $ is 
\begin{align}
    \begin{split}
        b_{1}<\frac{2\eta^{3}(T)}{3\sqrt{3}\sigma^{3}}\,.
    \end{split}
\end{align} 
Thus, for non-vanishing $b_{1}$, the approximate flat direction axes does not appear at $T_{C}$ but at
\begin{align}
    \begin{split}
        \frac{T}{T_{C}}
        =\sqrt{1-\frac{3\sqrt{3}}{2}b_{1}}\,.
    \end{split}
\end{align}
For $b_{1}>\frac{2}{3\sqrt{3}}\simeq 0.38$, the flat direction does not appear even at $T=0$, thus strings are never formed.}, as the curvature of the potential around the origin becomes negative.
To show this statement, let us use a polar decomposition defined by $\Phi = |\Phi|e^{i\theta}$, where $\theta$ is a collective coordinate.
The total effective potential can then be rewritten as
\begin{align}
    \mathcal{V}=\frac{1}{2}\lambda\left(|\Phi|^{2}-\eta^{2}(T)\right)^{2}+ 2\lambda b_{1} \sigma^{3}|\Phi|\cos{\theta}. \label{eq:sinne-Gordon potential}
\end{align}
In this basis, the original approximate $U(1)$ symmetry can be understood as an approximate shift symmetry $\theta \to \theta + \alpha$, with $\alpha$ being a real parameter.
This shift symmetry is slightly violated by non-vanishing $b_1$.
A flat direction is thus slightly lifted, leading to a  unique minimum of the potential at $\theta = \pi$.
Although there is no flat direction in a strict sense, it still accommodates cosmic string-like configuration when $\theta$ changes $2\pi n$ ($n= \pm 1,\pm 2, \cdots$) around a line in the position space.

\begin{figure}
    \centering
    \includegraphics[width=7cm]{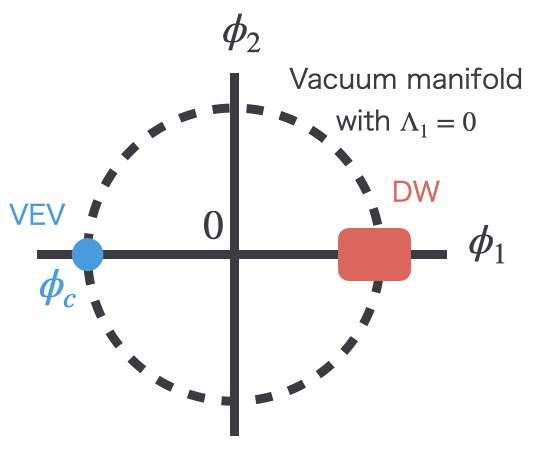}
    \caption{
    A contour of an approximate flat direction of the total effective potential  for $T<T_C$ in the case of $N = 1$.
    The black dashed circle represents the exact flat direction of the total effective potential in the absence of an explicit symmetry breaking, $b_1=0$. Once a non-vanishing but small $b_1$ turns on, the flat direction is slightly lifted, leading to the unique minimum indicated by the blue dot.
    When the phase of $\Phi=(\phi_1+i\phi_2)/\sqrt{2}$ changes by $2\pi$ around the origin, a domain wall is formed since the potential height around the red region is higher than that of the potential minimum (the blue dot).
    }
    \label{fig:conjecture-n1}
\end{figure}

Such a configuration is found in Ref.~\cite{Vilenkin:1982ks} by integrating out the massive mode $|\Phi|$ and solving the sine-Gordon equation for $\theta$.
It is known that this field configuration is a composite non-topological soliton in which a domain wall is attached to a cosmic string.
Fortunately, this feature can be understood from the shape of potential $\mathcal{V}$ given by \eqref{eq:sinne-Gordon potential} without solving the equation of motion.
When $\theta$ changes by $2\pi$ along a curve in the position space, there exists a domain wall at $\theta =0$ sourced by the cosine potential in $\mathcal{V}$.
Therefore, a single domain wall is necessarily attached to a cosmic string.
This feature is represented in Fig.~\ref{fig:conjecture-n1}.

Although an approximate flat direction permits a non-trivial solitonic configuration, 
it is not possible to realize such a configuration for a homogeneous $\Phi$.
However, inhomogeneous fluctuations induced by thermal plasma are always present.
If such thermal fluctuations overcome the potential height around $\Phi=0$, one may expect the formation of such a composite soliton.
We will discuss this possibility in Sec.~\ref{subsec:result n=1}.
The evolution of such a string-wall system after its  formation is investigated in the context of axion strings though the formation process could be rather different.
When (the inverse of) the Hubble parameter is comparable to the wall thickness, an effect of domain wall tension becomes important because it is not possible to sustain the cosmic string against the domain wall tension.
Such string-wall network is quickly  collapsed as is confirmed by numerical simulations~\cite{Hiramatsu:2010yn,Hiramatsu:2012gg}.

\subsection{The $N=2$ case}\label{subsec:setup-n2}

Let us proceed with the $N=2$ case. Here
the original $U(1)$ symmetry is explicitly broken down to $Z_2$.

With the paramaterization $\Phi =(\phi_1+i\phi_2)/\sqrt{2}$, the total effective potential can be rewritten as 
\begin{equation}
\begin{split}
    &\mathcal{V}
    =\frac{1}{2}\lambda\left(\frac{1}{4}(\phi_{1}^{2}+\phi_{2}^{2})^{2} +m_1^2(T)\phi_{1}^{2}+m_2^2(T)\phi_{2}^{2}\right),\\ \mbox{where\ \ }
    &m_1^2(T)= \frac{1}{3}T^2-(1-b_{2})\sigma^{2}\ \ \mbox{and\ \ } m_2^2(T) = \frac{1}{3}T^2-(1+b_{2})\sigma^{2}.
\end{split}
\end{equation}
Clearly, there exists an exact $Z_2$ symmetry whose transformation is defined by $\phi_1\to -\phi_1$ in addition to the $Z_2^{\rm CP}$ symmetry.
Thus, $\partial V/\partial \phi_1 = \partial V/\partial \phi_2=0$ is always satisfied at the origin 
$\phi_1=\phi_2=0$.

Let us assume the temperature is high enough initially, so that $m^2_2(T)>0$.
Since $m^2_1(T)>m_2^2(T)$, there is a unique minimum at the origin.
At temperature $T_{C_2}$ such that $m^2_2(T_{C_2})=0$ corresponding to $T_{C_2}=\sqrt{1+b_2}\,T_C$, a second-order phase transition with spontaneously broken $Z_2$ takes place.
Below $T_{C_2}$, the field $\phi_2$ condensates $\langle\phi_2\rangle \neq 0$ (with $\langle \phi_1 \rangle =0$), and the system is in the spontaneously broken $Z^{\rm CP}_2$ phase.
Thus, at least domain walls are formed, in contrast to the $N=1$ case, no matter how large is the explicit breaking of $U(1)$.
An additional characteristic temperature of the system is $T_{C_1}$, defined by $m^2_1(T_{C_1})=0$, corresponding to $T_{C_1}=\sqrt{1-b_2}\,T_C$.
Note that $T_{C_1}$ is the temperature below which the curvature of the total effective potential for $\phi_1$-direction becomes negative.
\footnote{If the explicit breaking is large ($b_2 > 1$), then $m^2_1(T)>0$ for any temperature and $T_{C_1}$ is ill-defined, implying strings are not formed.
In this case, the origin becomes the saddle point for $T<T_{C_2}$.
Since there are two disconnected potential minima, a domain wall is formed unless there is no bias to choose one of them.
Note that it is energetically favorable to generate a domain wall configuration with $\phi_1=0$. In this paper, we mainly focus on the case $0\leq b_{N}<1$, as already mentioned.}

\begin{figure}
    \centering
    \includegraphics[width=13cm]{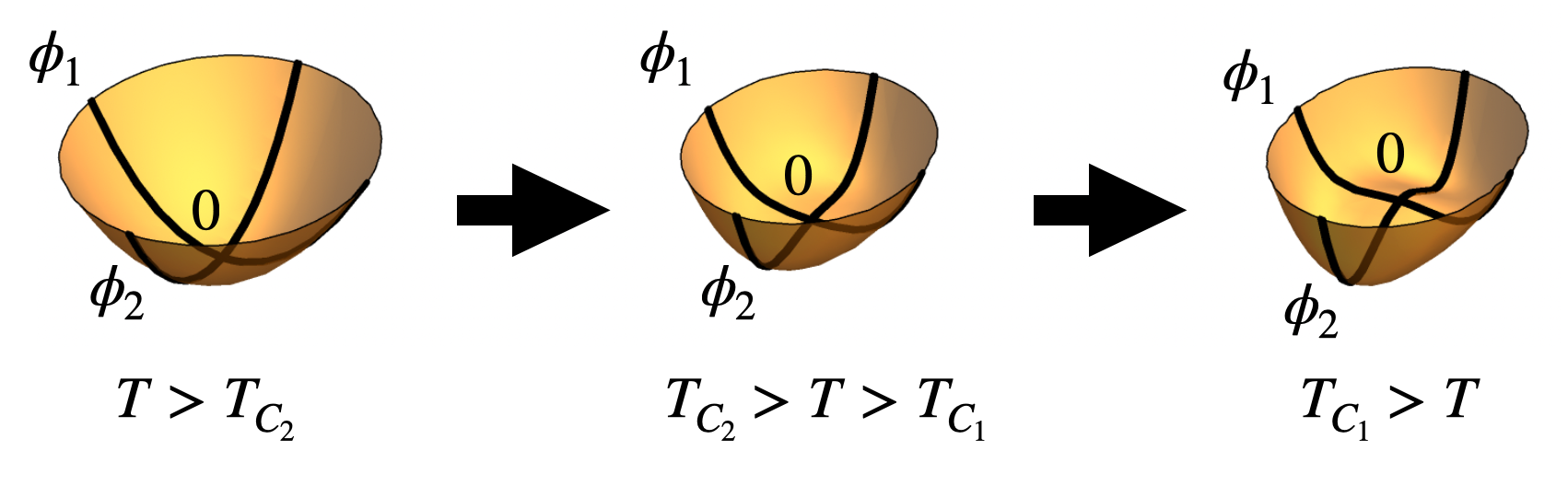}
    \caption{
    Total effective potentials with an explicit breaking $0<b_2<1$ on $(\phi_1,\phi_2)$-plane are shown for $T>T_{C_2}$ (the left), $T_{C_2}>T>T_{C_1}$ (the middle) and $T<T_{C_1}$ (the right) in the case of $N=2$.
    }
    \label{fig:potential-n2}
\end{figure}

\begin{figure}
    \centering
    \includegraphics[width=13cm]{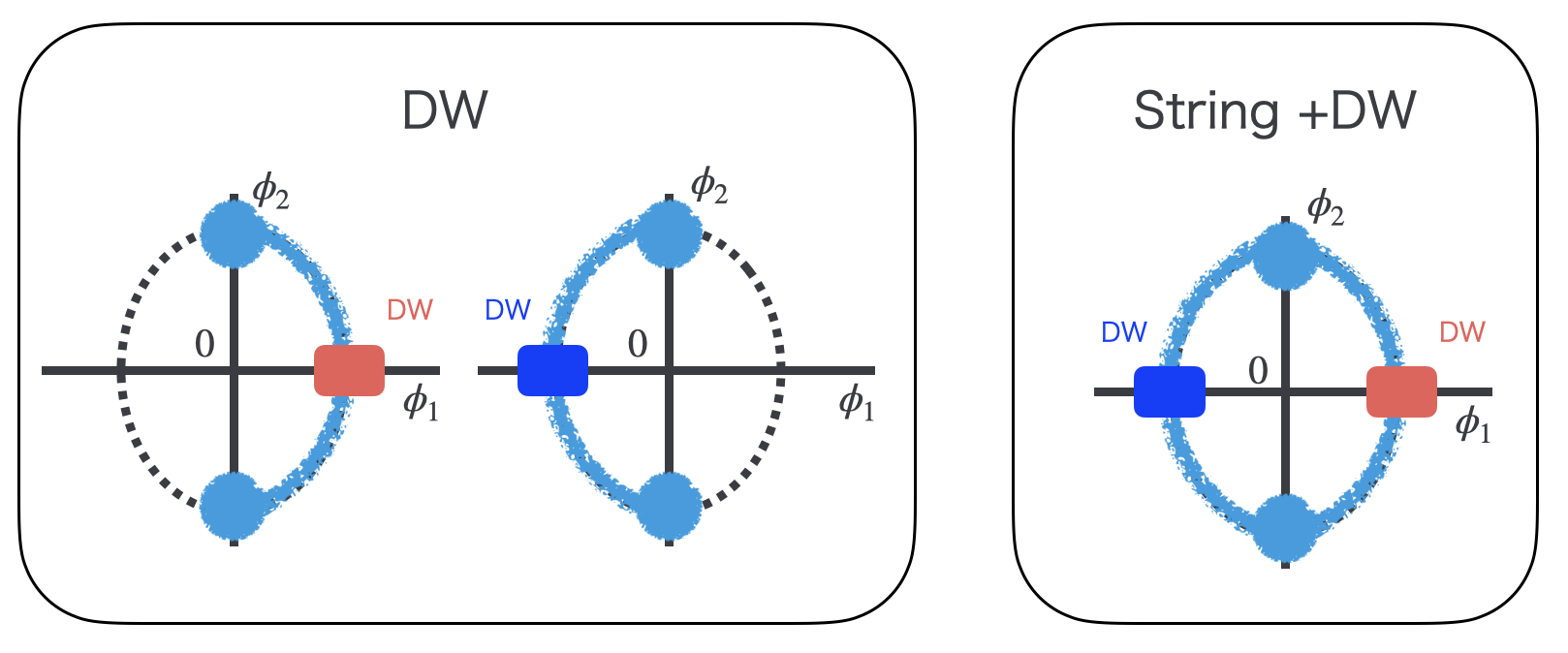}
    \caption{
    Possible defects formed in the case of $N=2$ are depicted for low temperature $T<T_{C_2}$ with a small explicit breaking $b_2<1$. The left panel shows two types of domain wall configurations that connect two disconnected minima by passing $\phi_1>0$ region (the red color) and $\phi_1<0$ region (the blue color).
    The right panel shows a line-like topological defect which can be formed when a phase of $\Phi$ changes $2\pi$ around a line in the position space.
    In this case two domain walls are necessarily attached to the cosmic string because there exist two potential walls at $\theta = 0$ and $\pi$.
    }
    \label{fig:conjecture-n2}
\end{figure}

Figure \ref{fig:potential-n2} shows the shape of the total effective potential.
At high temperatures $T>T_{C_2}$, there is a unique minimum at the origin.
For intermediate temperatures $T_{C_2}>T>T_{C_1}$, the origin becomes a saddle-point and $Z_2$ symmetry is spontaneously broken.
A possible defect in this case is a domain wall with a configuration similar to the $b_2>1$ case.
Two disconnected minima are smoothly connected by a domain wall configuration with $\phi_1=0$ in order to minimize the potential energy.
For $T<T_{C_1}$, the curvature of $\phi_1$-direction at the origin is negative.
In this region, it is useful to rewrite the total effective potential in the polar coordinate $\Phi =|\Phi|e^{i\theta}$, as
\begin{align}
    \mathcal{V}
    =\frac{1}{2}\lambda\left(|\Phi|^{2}-\eta^{2}(T)\right)^{2}+ \lambda b_{2} \sigma^{2}|\Phi|^{2}\cos{2\theta}.
\end{align}
Similarly to the $N=1$ case, the flat direction parameterized by $\theta$ is slightly lifted by the explicit breaking, $b_2$.
An approximate flat direction admits cosmic string-like configuration when the phase of $\Phi$ changes by $2\pi$  in position space.
Since there exists two potential walls at $\theta=0$ and $\theta = \pi$, the cosmic string is necessarily attached to {\it two} domain walls, as shown in Fig.~\ref{fig:conjecture-n2}. Hereafter, the one at $\theta=0$ will be referred to as DW-A and the other one as DW-B.

The evolution of string-wall network is investigated in Ref.~\cite{Kawasaki:2014sqa}.
Contrary to the $N=1$ case, a string-wall network is stable as observed in our simulation. This could be interpreted as the result of strings are connected by domain walls whose tensions forbid strings from decaying.
Since the domain wall is a surface-like topological defect, its energy density decreases slower than those of relativistic and non-relativistic matter. 
Hence the energy density of the domain wall quickly dominates, which could be in tension with cosmological observations~\cite{Zeldovich:1974uw}.

A remarkable feature in the $N=2$ case is that cosmic strings may be produced as long as $0<b_2<1$.
For $T<T_{C_1}$, there can be two types of domain walls.
One that connects two disconnected minima by crossing the $\phi_1>0$ region, and another one that connects them through the $\phi_1 < 0$ region. 
When $Z_2$ symmetry is an exact symmetry, two minima are randomly distributed in the early Universe at least beyond the horizon scale (which is the inverse of the Hubble parameter at $T=T_{C_2}$) because causality forbids such a long-range correlation.
Therefore two types of domain walls are randomly generated.
Then the cosmic string attached to two domain walls may be inevitably produced below $T_{C_1}$ in order to smoothly connect two disconnected minima.
Clearly, this production mechanism strongly relies on the presence of $Z_2$ symmetry which is absent in the case of $b_1 \neq 0$ ($N=1)$. 

Finally, in the case that $b_1 \neq 0$ and $b_2 \neq 0$ simultaneously, the potential has a local well at $\theta=\pi$ due to $b_{1}\neq 0$ and two local walls at $\theta=\pi/2$ and $\theta=3\pi/2$ due to $b_{2}\neq 0$ below $T=T_{C_{2}}$. The depths of those wells depend on the size of $b_{1}$ and $b_{2}$.

In the case of $b_{1}\gtrsim b_{2}$\footnote{The precise expression of the condition in which the local potential wall induced by $b_{1}\neq 0$ becomes deeper than that by $b_{2}\neq 0$ is,
\begin{align}
    \begin{split}
        b_{2}\leq \frac{1}{2}\left(-1 + \frac{T^{2}}{T_{C}^{2}} + \sqrt{-\frac{4\sqrt{2}b_{1}\langle \phi_{1}\rangle}{\sigma}+\left(1-\frac{T^{2}}{T_{C}^{2}}\right)^{2}}\right).
    \end{split}
\end{align} Note that this condition also depends on the cosmic temperature. Using Eq.~\eqref{eq:<phi_1> for T<T_C}, it reduces to
\begin{align}
    \begin{split}
        b_2\leq \frac{2b_{1}}{\sqrt{1-\frac{T^{2}}{T_{C}^{2}}}} +\mathcal{O}(b_1^{2}).
    \end{split}
\end{align} 
We denote this condition by $b_{1}\gtrsim b_2$, assuming that $2/\sqrt{1-T^2/T_{C}^2}=\mathcal{O}(1)$. 
In the zero temperature limit $T = 0$, the condition can be written as 
\begin{align}
    \begin{split}
        b_{2}\leq \frac{1}{2}\left(-1 + \sqrt{1+\frac{2^{8/3}b_{1}(2^{2/3} + (3\sqrt{3}b_{1} + \sqrt{-4+27 b_{1}^{2}})^{2/3}}{\sqrt{3}(3\sqrt{3}b_{1} + \sqrt{-4+27 b_{1}^{2}})^{1/3}}}\right).
    \end{split}
\end{align}
This exact expression at $T=0$ agrees with the approximate expression obtained from Eq.~\eqref{eq:<phi_1> for T<T_C} with more than $99\,\%$ accuracy for $b_{1}=10^{-1}$.
}
, the effective potential minimum induced by $b_1\neq 0$ is the global minimum and exists at any temperature.
Hence the situation is the same as with $b_1\neq 0$ and $b_2= 0$.

In the case of $b_{2}\gtrsim b_{1}$, two minima induced by $b_{2}\neq 0$ are deeper and are approximately degenerated.
Hence a composite soliton where two domain walls are attached to a single cosmic string is formed because of the $Z_2$ degeneracy which is similar to the case of $b_1=0$ and $b_2\neq 0$.
However, a crucial difference is that this $Z_2$ degeneracy is lifted by $b_1\ne 0$.
Consequently, two domain walls are not stable and are eventually collapse due to a pressure difference~\cite{Vilenkin:1981zs,Gelmini:1988sf,Larsson:1996sp}.

\subsection{The large $N$ case}
Let us discuss the $N=3$ and $N=4$ cases. For both cases, the original $U(1)$ symmetry is explicitly broken down to $Z_{N}$. Therefore, $\partial V/\partial\phi_{1}=\partial V /\partial \phi_{2}=0$ is always satisfied at the origin $\phi_{1}=\phi_{2}=0$, similarly to the $N=2$ case.

For $N=3$ and $N=4$, the potential is modified by the cubic and quartic terms of $\Phi$, respectively. Therefore, unlike the $N=2$ case, the temperature dependent mass $m^{2}(T)\equiv -\lambda\eta^{2}(T)$ is not modified by $b_{N}\neq 0$, i.e. a potential hill appears at the origin below $T=T_{C}$. 

In the $N=3$ case, the potential has three vacua at $\theta=\pi/3,\pi,5\pi/3$ below $T=T_{C}$. Therefore, at least, domain walls are formed no matter how large the explicit breaking of $U(1)$ is, as long as the initial temperature of universe is higher than $T_{C}$, and dominate the universe soon after their formation. As in the  $N=2$ case, an approximate flat direction admits cosmic string-like configuration when the phase of $\Phi$ changes $2\pi$ around a string in the position space, with three walls are attached.

For $N=4$,  the potential becomes unbounded for $b_{4}>1$. Therefore, we only consider $0<b_{4}<1$. As discussed previously, the domain walls are formed below $T=T_{C}$ as long as the initial temperature of the universe is higher, and they dominate the universe soon after their formation. A cosmic string is also formed at the position around which the phase of $\Phi$ changes by $2\pi$, with four domain walls attached to it.

\subsection{Effects of inhomogeneous thermal fluctuations}\label{sec:thermal fluctuation}

So far, we have neglected effects of inhomogeneous thermal fluctuations.
However, these effects play an important role to form topological defects in the case of $N=1$ as discussed in the Sec.~\ref{subsec:setup-n1}.
We will now include these effects in the analysis of defects formation.

Let us use the decomposition, $\Phi =(\phi_1+i\phi_2)/\sqrt{2}$, in the following analysis.
We take into account inhomogeneous fluctuations of $\phi_{i}$ and $\dot{\phi}_{i}$ such that two-point functions are given by free scalar fields in finite-temperature field theory,
\begin{align}
    &\langle\delta {\phi}_{i}(\bm{x},t)\delta {\phi}_{j}(\bm{y},t)\rangle_{\beta}=\delta_{ij}\int \frac{\mathrm{d}^{3}p}{(2\pi)^{3}{\omega_{p}}_{i}}n({\omega_{p}}_{i})e^{i\bm{p}\cdot (\bm{x}-\bm{y})},\nonumber\\
    &\langle\delta \dot{\phi}_{i}(\bm{x},t)\delta \dot{\phi}_{j}(\bm{y},t)\rangle_{\beta}=\delta_{ij}\int \frac{\mathrm{d}^{3}p}{(2\pi)^{3}}{\omega_{p}}_{i}n({\omega_{p}}_{i})e^{i\bm{p}\cdot(\bm{x}-\bm{y})},\label{eq:initial configuration}\\
    &\langle\delta {\phi}_{i}(\bm{x},t)\delta \dot{\phi}_{j}(\bm{y},t)\rangle_{\beta}=0,\nonumber
\end{align}
In this expression, $\delta {\phi}_{i}\equiv \phi_{i}-\langle {\phi_{i}}\rangle$, where $\langle {\phi_{i}}\rangle$ is the position of the global minimum of the potential, $\mathcal{V}$.
Here, $\beta\equiv 1/T$, and $n(\omega_i)$ is the Bose distribution function in the absence of chemical potential given by
\begin{align}
    n(\omega_i)=\frac{1}{e^{\beta\omega_i}-1},
\end{align}
where $\omega_i\equiv \sqrt{|\bm{p}|^2+m_i^2}$.
Note that the zero-temperature contributions of two-point functions are subtracted because they lead to the divergence caused by the short distance singularity,
$m_i^2$ is defined as the temperature-corrected mass term at the potential minimum,
\begin{align}
m_{i}^{2}=\frac{\partial^{2}\mathcal{V}}{\partial\phi_{i}^{2}}(\phi_{i}={\langle \phi_{i}}\rangle).\label{eq:mass}
\end{align}
For simplicity, we will neglect the second term of Eq.~\eqref{eq:initial configuration}, but we will include it in the numerical simulation.
Then one can roughly estimate the root-mean-square of amplitudes of scalar fields caused by thermal fluctuations as follows.
The first line of the Eq.~\eqref{eq:initial configuration} can be expressed as
\begin{equation}
\begin{split}
    &\langle\delta{\phi}_i(\bm{x},t)\delta{\phi}_j(\bm{y},t) \rangle_\beta = \delta_{ij}\dfrac{T^2}{2\pi^2}G\left(\dfrac{m}{T},|\bm{x}-\bm{y}|T\right),\\
    &G(z_1,z_2)\equiv \int \mathrm{d}x\,\left(\frac{x^2}{\sqrt{x^2+z_1^2}}\frac{1}{e^{\sqrt{x^2+z_1^2}}-1}\frac{\sin\left(xz_2\right)}{xz_2}\right). \label{eq:thermal fluctuations}
\end{split}
\end{equation}
Owing to the Boltzmann suppression, this integral converges even in the short-distance limit, $\bm{x}=\bm{y}$.
In the massless and short-distance limits, one obtains root-mean-square of the amplitude of $\phi_i$,
\begin{align}
    \sqrt{\left\langle \delta{\phi}^2_i(t) \right\rangle_\beta} =\dfrac{T}{2\sqrt{3}}, ~\left(\dfrac{m_i}{T}\to 0\right). \label{eq:the root-mean-square of amplitude}
\end{align}
A non-zero $m_i/T$ suppresses $\sqrt{\langle\delta\phi_i^2(t)\rangle_\beta}$ by a factor of order one because
$m_i$ in our case is roughly given by $m_i\sim \sqrt{\lambda}(T- \sigma)\lesssim \sqrt{\lambda}T$ with a small correction from an explicit breaking term, which justifies $m_i/T<1$ for $\lambda <1$.
We find that the root-mean-square of amplitudes of scalar fields are roughly given by the temperature.

We shall discuss the time evolution of mean amplitudes of field fluctuations to clarify the formation of defects including the effect of thermal fluctuation based on the effective potential given by Eq.~\eqref{eq:potential}.
At high-temperatures, the root-mean-square of amplitudes of $\phi_i$ becomes large according  to Eq.~\eqref{eq:the root-mean-square of amplitude}.
In an expanding universe whose metric is given by Eq.~\eqref{eq:metric}, the amplitudes of scalar field oscillations with the potential Eq.~\eqref{eq:potential} decrease by cosmic expansion.
When the amplitude of a scalar field is large enough to oscillate with a quartic potential, it decreases in proportion to $a^{-1}$.
In the opposite case where the amplitude of a scalar field is small enough to oscillate with a quadratic potential, in general, it decreases in proportion to $a^{-3/2}$ if the mass term is constant with respect to cosmic time.
However, in our set up, the mass of the scalar field is roughly given by $\sqrt{\lambda} T$ for $T>T_C$, and hence, this mass term also gets smaller through cosmic expansion.
In this case,  the amplitude of the scalar field decreases in proportion to $a^{-1}$.
The detailed derivation of this dependence is summarized in App.~\ref{app:scalar field oscillation}.
Therefore, in either case, the amplitude of the scalar field oscillation decreases in proportion to $a^{-1}$.
Hence the root-mean-square of amplitudes of $\phi_i$ generated by thermal fluctuations at $t= t_1$ becomes
\begin{align}
    \left(\frac{a(t_1)}{a(t_2)}\right)\sqrt{\langle \delta \phi_i^2(t_1)\rangle_{\beta_1}}\sim T_1\left(\frac{T_2}{T_1}\right)\sim T_2\sim \sqrt{\langle\delta\phi_i^2(t_2)\rangle_{\beta_2}}, \label{eq:initial temperature dependence} 
\end{align}
at $t=t_2>t_1$.\footnote{Here, we have neglected the possible change of the effective number of entropy
degrees of freedom. If we take such a change into account, the quantity in the far left hand side is suppressed by (power of one-third of) the change of the effective number of entropy degrees of freedom and be smaller than $T_2$.}
Here, $T_{1,2}$ are cosmic temperatures at $t=t_{1,2}$, respectively, while we use approximations, $\sqrt{\langle \delta \phi_i^2(t)}\rangle_\beta \sim T$.
The root-mean-square of amplitudes of $\phi_i$ generated at $t=t_2$ is given by $\sqrt{\langle \delta \phi_i^2(t_2)\rangle_\beta}\sim T_2$.
Therefore, we the typical amplitude of the fluctuations generated at higher temperatures are comparable to that generated. Thus, it is sufficient to consider fluctuations generated at $T\sim T_C$.

For $N=1$, defect formation strongly depends on the size of fluctuations since $\langle \phi_1(T)\rangle\neq 0$ at any temperature.
Since an approximate flat direction with non-trivial topology $\mathbb{S}^1$ emerges at $T=T_C$ (which is defined as the critical temperature in the limit of $b_1\to 0$ as discussed in Sec.~\ref{subsec:setup-n1}), the cosmic string formation takes place when the fluctuation is large enough at this temperature.\footnote{Even for $T<T_C$, cosmic string formation may take place by a classical motion of $\phi_i$ which overcomes the saddle-point (top of the hill) of the effective potential. However, the probability of this sphaleron-like process is proportional to the Boltzmann factor $e^{-\mathcal{V}(\phi_i=0,T)/T}$~\cite{Affleck:1980ac}, and thus, one may safely neglect this contribution for $T<T_C$.}
Assuming $b_1 \ll 1$, the cosmic string formation may take place when
\begin{align}
\sqrt{\langle \delta \phi_1^2(T_C)\rangle_\beta} \gtrsim |\langle \phi_1(T_C)\rangle|. \label{eq:string formation criterion}
\end{align}
It is clear that assuming $\sqrt{\langle \delta\phi_1^2(t)\rangle _\beta}=0$ as in Sec.~\ref{subsec:setup-n1}, no cosmic string formation takes place.
Also, when $\langle \phi_1(T_C)\rangle =0$, which is the case of the absence of explicit breaking term, the above condition is satisfied for any non-zero $\langle \delta \phi_i^2(T_C)\rangle_\beta\neq 0$.
The above condition can be further translated into the bound on the amount of the explicit breaking from Eqs.~\eqref{eq:the root-mean-square of amplitude} and \eqref{eq:minimum at the critial temperature}.
The result is given by
\begin{align}
    b_1 \lesssim \dfrac{1}{16\sqrt{2}}\simeq 0.04.\label{eq:critical value}
\end{align}

This result Eq.~\eqref{eq:critical value} strongly relies on the shape of the one-loop effective potential $\mathcal{V}$ and inhomogeneous fluctuations approximated by free field theory given in Eq.~\eqref{eq:initial configuration}.
However, to be strict, it is known that one-loop effective potential and the free field approximation cannot be justified around $T\simeq T_C$ for $b_1=0$ due to a non-perturbatively large fluctuation of the order parameter called critical fluctuation~\cite{Weinberg:1974hy}.
Although an inclusion of this effect on formation of defects is very challenging and beyond the scope of the present paper, one may make qualitative arguments and give at least the lower limit of the condition on the defect formation even without taking such non-perturbative effects into account.

\subsection{Explicit symmetry breaking effects from the QCD instanton and wormholes}\label{sec:explicit breaking effects}

Let us discuss explicit breaking effects arising from the QCD instanton and creation of wormholes.
For simplicity, we only discuss the case with $N=1$.

We first focus on the invisible axion models, where it is commonly assumed that $b_1$ is only sourced by the QCD instanton effect with the high breaking scale, $\sigma\gtrsim 10^{9}\,{\rm GeV}$.
Generically, effects of QCD instantons are non-perturbative, and thus, it is challenging to estimate $b_1$ without non-perturbative numerical computations.
However, the running coupling constant of the color gauge group is perturbative in the high temperature region at least for $T>T_C\sim \sigma \gtrsim 10^{9}\,{\rm GeV}$ owing to the asymptotic freedom.
Therefore one may safely perform semi-classical approximation to estimate reasonable value of $b_1$ as follows.
Neglecting the interaction of instantons (called the dilute instanton gas approximation)~\cite{Callan:1977gz}, $b_1$ can be expressed by the integration of instanton size represented by $\rho$ as 
\begin{align}
	b_1 (T)=\dfrac{1}{\lambda\sigma^3|\Phi(T)|}\int_{1/\Lambda_{\rm UV}}^{1/\Lambda_{\rm SM}}\mathrm{d}\rho \,n(\rho,T); \label{eq:integration of instanton density}
\end{align}
 $\Lambda_{\rm SM}\sim 100\,{\rm GeV}$, $\Lambda_{\rm UV}\sim \sigma$, $n(\rho)$ are IR and UV cutoff of the theory and the instanton density, respectively. The quantity
$n(\rho)$ can be estimated by computing the instanton action and its fluctuation.
In the calculation of $n(\rho)$, suppressions from fermion zero-modes on the instanton background should be included.
Assuming no new colored particles in addition to the SM quarks and gluons up to an energy scale $T_C$, $n(\rho,T)$ can be evaluated in the finite temperature field theory as~\cite{Gross:1980br,Flynn:1987rs,Dine:1986bg}
\begin{align}
	n(\rho,T)=   \det \left(\dfrac{y_u}{4\pi}\right) \det\left(\dfrac{y_d}{4\pi}\right)\dfrac{1}{\rho^5}K(g_s,\rho T) \exp\left[-\frac{8\pi^2}{g_s^2}-\frac{(\pi\rho T)^2}{3}(2N_C+N_F)\right],
\end{align}
where $N_C = 3,~N_F =6$ and $g_s$ is the strong gauge coupling constant evaluated at an energy scale $\rho^{-1}$. The numerical coefficient
$K(g_s, \rho T)$ is less sensitive to $g_s$ and $\rho T$ and for our analysis we will  approximate $K(g_s,T\rho)\sim 1$; a precise expression of $K(g_s, \rho T)$ can be found in Ref.~\cite{Gross:1980br}. Finally,
$y_u$ and $y_d$ are $3\times 3$ SM up and down-type Yukawa couplings with the Higgs field, respectively.\footnote{Factors of determinant of the SM Higgs-Yukawa coupling can be understood as the presence of  approximate quark flavor symmetries because $b_1$ must be invariant under $y_u \to U_u y_u V_u$ and $y_d \to U_d y_d V_d$ where $U_{u,d},~V_{u,d}\in SU(3)$.}
Putting the one-loop beta function of $g_s$ into the above expression, one obtains
\begin{align}
	n(\rho,T)\sim  \det \left(\dfrac{y_u}{4\pi}\right) \det\left(\dfrac{y_d}{4\pi}\right)\dfrac{1}{\rho^5}\left(\rho\Lambda_{\rm QCD}\right)^{(11N_C-2N_F)/3} \exp\left[-\frac{(\pi\rho T)^2}{3}(2N_C+N_F)\right],
\end{align}
where $\Lambda_{\rm QCD}\sim 100\,{\rm MeV}$ is the QCD confinement scale.
If there exist new colored fields for an intermediate energy scale $\Lambda_{\rm SM}< \Lambda <\Lambda_{\rm UV}$, their effect on the beta function of $g_s$ must be included.
The integration Eq.~\eqref{eq:integration of instanton density} is dominated at $\rho\sim  T^{-1}$, where $T$ plays a role of IR cutoff of the instanton size for $T\gg \Lambda_{\rm QCD}$.
This is because thermal field theories have a periodicity for the temporal direction with period $T^{-1}$ and hence the instanton size has to be shorter than this scale.
Using the observed values of Yukawa couplings, $|\Phi(T)|\sim \sigma$ and $T\sim T_C\sim \sigma$, we finally arrive at the following expression,
\begin{align}
	b_1(T_C)\sim 10^{-23}\dfrac{1}{\lambda}\left(\dfrac{\Lambda_{\rm QCD}}{T_C}\right)^7\sim \frac{1}{\lambda}10^{-93};
\end{align}
where we have used $T_C\sim 10^{9}\,{\rm GeV}$.
This value is many order magnitude smaller than that of the critical value, Eq.~\eqref{eq:critical value}, because of the strong suppression from the running of $g_s$ at high-temperatures.
Therefore (the invisible QCD) axion string is safely formed if a global $U(1)$ symmetry is violated only by the QCD instanton effect.

We next discuss the explicit breaking effects arising from creation of wormholes;
detailed discussion on this topic is reviewed in e.g.~Ref.~\cite{Hebecker:2018ofv}.
The simplest wormhole solution in Einstein gravity with a single three form field which is dual of the shift symmetric scalar field is the Giddings-Strominger wormhole~\cite{Giddings:1987cg}.
When explicit breaking arises from such a wormhole formation, the amount of the explicit symmetry breaking can be estimated by the semi-classical approximation as
\begin{align}
	b_1 \sim \left(\frac{M_{\rm Pl}}{\sigma}\right)^4e^{-S_E},
\end{align}
where $M_{\rm Pl}$ and $S_E\sim M_{\rm Pl}/\sigma$ are the Planck mass and the wormhole action, respectively.
Note that $S_E \gg \mathcal{O}(1)$ is required to justify the semi-classical approximation.
(The wormhole solution of a Goldstone model with minimally coupled Einstein gravity can be found in Ref.~\cite{Kallosh:1995hi}.)
A characteristic feature of the explicit breaking induced by wormhole formation is that $b_1$ is independent of ambient temperature, contrary to the case of the QCD instanton effect.
Since we are interested in the situation where the explicit breaking is induced mainly by quantum gravity effects rather than by the QCD instanton, we treat $b_1$ as a constant free parameter.

\section{Setup of numerical simulations}
\label{sec:numerical simulation}

Let us discuss the setup of our numerical simulations.

We decompose the complex scalar field as $\Phi = (\phi_1+i\phi_2)/\sqrt{2}$.
The equations of motion for the scalar fields are given by
\begin{align}
    \ddot{\phi}_{i}+3H(t)\dot{\phi}_{i}-\frac{1}{a(t)^{2}}\nabla^{2}\phi_{i}+\frac{\partial \mathcal{V}(\Phi,T)}{\partial \phi_{i}}=0,~(i=1,2)\label{eq:eom}
\end{align}
where an $\dot{a}$ represents the derivative of $a$ with respect to the cosmic time, $t$. Our numerical calculations are based on the staggered leapfrog method with second order accuracy both in time and in space. In the radiation-dominated Universe, the Hubble parameter $H(t)$ and $t$ are given by
\begin{align}
    H^{2}(t)
    =\left(\dfrac{\dot{a}}{a}\right)^2=\frac{8\pi}{3M_{\rm pl}^{2}}\frac{\pi^{2}g_{*}(t)T^{4}}{30},\quad t=\frac{1}{2H(t)}\equiv \frac{\epsilon}{T^{2}},
\end{align}
where $M_{\rm pl}\simeq 1.2\times 10^{19}\,{\rm GeV}$ is the Plank mass and $g_{*}$ is the total number of degrees of freedom for the relativistic particles. We define a dimensionless parameter $\zeta$ as
\begin{align}
    \zeta=\frac{\epsilon}{\sigma}=\left(\frac{45 M_{\rm pl}^{2}}{16\pi^{3}g_{*}\sigma^{2}}\right)^{\frac{1}{2}}.\label{eq:hubble}
\end{align}

We prepare a cubic lattice to perform the classical lattice simulation.
To obtain reliable results, finite size box and non-zero lattice size effects must be suppressed.
Hence, the lattice size and length of the box must be chosen such that the physical size of the whole box at final time is comparable to or larger than the Hubble scale, and the physical size of each lattice at final time is comparable to or shorter than the typical width of defects.
In the presence of an explicit symmetry breaking term, cosmic strings attached to domain walls appear in our simulation, and thus, there are two typical length scales corresponding to cosmic string and domain wall thickness.
Since we consider only the case where the explicit breaking term is smaller than that of the symmetric one, $0\leq b_{N}<1$, the typical width of domain walls is comparable to or longer than that of cosmic strings.
Therefore, the lattice size is determined by the cosmic string thickness.
As a result, the box size and a dynamical range of the simulation must be chosen such that following relations are satisfied.
\begin{equation}
    \Delta \lesssim \dfrac{1}{\sqrt{\lambda}\sigma},~H(t_{\rm f})^{-1}\gtrsim N_l^{1/3}\Delta  ;\label{eq:simulation requirement}
\end{equation}
$t_{\rm f},~N_l$ and $\Delta$ are the final time of the simulation, the total number of cubic lattice and the comoving lattice size, respectively.
In this calculation, we set the string thickness as $\delta =1/(\sqrt{\lambda} \sigma)$.
We determine the time-step $\delta t$ as $\delta t=\Delta \times a(T_{\rm in})/a(T_{C}) /100$, that is, $1\%$ of the initial physical lattice size, where $T_{\rm in}$ is the initial temperature of the simulation.
We numerically generate a a configuration of $\phi_i$ and $\dot{\phi}_i$ at $T=T_{\rm in}$ in momentum space and use Fourier transformation to obtain position space field configurations given by Eq.~\eqref{eq:initial configuration}.
In order to discuss the initial temperature dependence $T_{\rm in}$, we treat it as a free parameter and simulate with several initial temperatures.

We identify strings and estimate their length following the procedure given in Ref.~\cite{Yamaguchi:2002sh}. For DWs, we determine the position of the points where DW-A (B) intersects lattices, and estimate the area of each wall.

The energy densities of strings and DWs are paramaterized by 
\begin{align}
    \rho_{\rm string} = \xi_{\rm string}(t) \dfrac{\mu}{t^2},\quad \rho_{\rm DW\text{-}{A,B}} = \xi_{\rm DW\text{-}A,B}(t)\dfrac{s}{t},
\end{align}
where $\mu$ and $s$ are string and DW tensions, respectively.
In the numerical simulation, $\xi_{\rm string, DW}(t)$ can be obtained by
\begin{align}
    \xi_{\rm string}(t)=\frac{lt^{2}}{N_la(t)^{3}\Delta^{3}},\quad \xi_{\rm DW\text{-}A,B}(t) = \frac{A_{\rm A,B}t}{N_{l} a(t)^{3}\Delta^{3}}
\end{align}
where $l,\,A_{\rm A,B}$ are the total physical length of strings and areas of DWs A and B, respectively.
We will show the arithmetic mean value of $\xi(t)$ for $10$ different initial configurations with given values of $T_{\rm in}$ in the next section.

We perform the computation with $N_l = 128^{3}$ of lattices.
We use fixed parameters $\lambda=0.08$ and $\zeta = 10$.
The comoving lattice size and the final cosmic time of simulation are set to be $\Delta = 1/(2\sqrt{\lambda}\sigma)$ and $t_{\rm f}=N_l^{1/3}\Delta /4$.
Since $\zeta=10$ corresponds to $\sigma \sim 10^{17}\,{\rm GeV}$ for $g_*=100$, this parameter choice is unrealistic because we consider the case in which production of topological defects takes place after inflation.
However, $N_l \sim (\sqrt{\lambda}\zeta \sigma^2/T_{\rm f}^2)^3$, where $T_{\rm f}$ is the temperature at $t_{\rm f}$, and hence, a larger $\zeta>10$ requires a larger $N_l$ which scales as $N_l\propto \zeta^3$.
Unfortunately, it is very hard to simulate with a large value of $\zeta =10^2$.
This simply reflects the fact that it is difficult to make a large hierarchy between the Planck scale and the breaking scale $\sigma$ in the practical computation.

In summary, we fix other parameters including lattice separation as 
\begin{align}
	\zeta \equiv \left(\frac{45M_{\rm Pl}^2}{16\pi^3g_*\sigma^2}\right)^{\frac{1}{2}} = 10,~N_l = 128^3,~\lambda =0.08,
\end{align}
and there are following free parameters
\begin{align}
 \{b_{N},~T_{\rm in}\}.
\end{align}
In App.~\ref{app:simluation with a larger box size}, we also present the results of the simulation of a larger box size with $N_{l}=256^{3}$ by fixing $T_{\rm in}=2T_{C}$ to investigate the finite box size effects.


\section{Results of numerical simulations}\label{sec:result}

In what follows we first discuss our results in the case that there is no explicit breaking of a global $U(1)$ symmetry ($b_{N}=0$) in Sec.~\ref{subsec:result SSB}.
Then, we present our results for $0<b_1 <1, ~b_2 = 0$ and $b_1 = 0,~0<b_2 < 1$ in Sec.~\ref{subsec:result n=1} and Sec.~\ref{subsec:result n=2} and various initial temperatures $T_{\rm in}$.

\begin{figure}
    \centering
    \includegraphics[width=13cm]{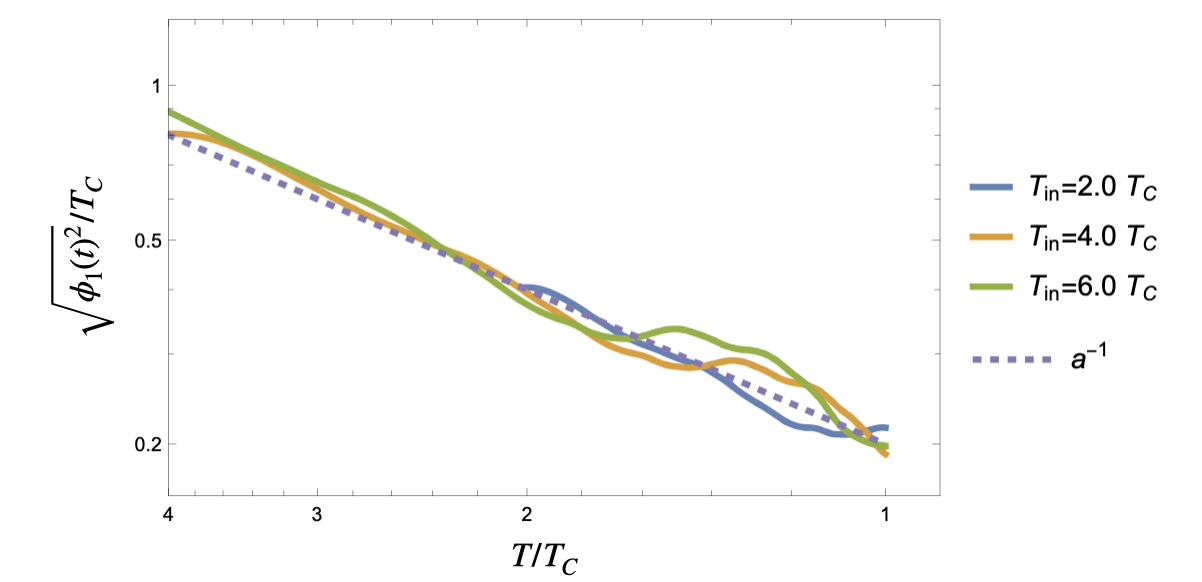}
    \caption{The evolution of $\langle\sqrt{\phi_{1}(t)^{2}}\rangle$ for $T_{\rm in}=2.0T_{C}$ (blue solid curve), $T_{\rm in}=4.0T_{C}$ (yellow solid curve) and $T_{\rm in}=2.0T_{C}$ (green solid curve) during $T_{C}<T<T_{\rm in}$. The values of $\langle \sqrt{\phi_{2}(t)^{2}}\rangle $ are almost the same and we omit to plot. The purple dashed curve shows the behavior $a^{-1}$.
    }
    \label{fig:Amplitude}
\end{figure}

Let us first discuss the effect of $T_{\rm in}$.
According to our analytic estimate, Eq.~\eqref{eq:initial temperature dependence}, the typical amplitudes of initial configurations generated at high-temperature $T\gg T_{C}$ decay in proportion to $a^{-1}$ and are comparable to those of the configurations generated at $T\sim T_{C}$.
To confirm this statement, we numerically follow the time evolution of the mean amplitudes of the scalar field.
Figure \ref{fig:Amplitude} shows the time evolution of the mean amplitude of $\phi_1$ generated at three different initial temperatures.
The behavior of $\phi_2$ is almost the same, and thus, we omit it for simplicity.
We see that all mean amplitudes are almost the same and are independent of the initial temperatures.
Furthermore, the mean amplitudes decay in proportion to $a^{-1}$, which is in agreement with our analytic study.

\subsection{Case without explicit breaking operators}\label{subsec:result SSB}

We will discuss numerical results for $b_{1}=b_{2}=0$, corresponding to the case where the global $U(1)$ symmetry is an exact symmetry.

\begin{figure}
    \centering
    \includegraphics[width=13cm]{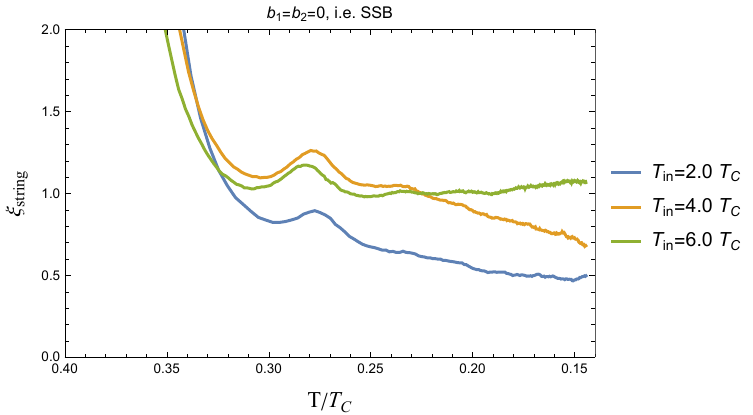}
    \caption{The time evolution of $\xi_{\rm string}$ for $T_{\rm in}=2.0T_{C}$ (blue), $T_{\rm in}=4.0T_{C}$ (yellow) and $T_{\rm in}=6.0T_{C}$ (green).}
    \label{fig:result_ssb}
\end{figure}

Figure~\ref{fig:result_ssb} shows the evolution of the string scaling parameter $\xi_{\rm string}$ for various initial temperatures.
For $T_{\rm in}=2T_C$, our setup is the same as that of Ref.~\cite{Yamaguchi:2002sh}.
We confirm that the evolution of the scaling parameter is qualitatively in agreement with the previous study.
For any initial temperatures we tried, $\xi_{\rm string}$ does not vanish at the final time of the simulation $T\simeq 0.15T_C$, when the volume of the simulation box is the same as the Hubble volume.
However, the finite box size effect makes $\xi_{\rm string}$ smaller at $T \simeq 0.15T_C$ as discussed in Ref.~\cite{Yamaguchi:2002sh}.
Although we cannot follow the long-term dynamics of the cosmic string in our simulation setup in the region $T<0.15 T_C$, it is sufficient to quantify the effect of the explicit symmetry breaking on $\xi_{\rm string}$ as we will see later.

\subsection{The $N=1$ case}\label{subsec:result n=1}

We present numerical results for $b_1 \neq 0$ and $b_2=0$ and  then discuss the cosmic string formation by comparing results obtained previously.

\begin{figure}
    \centering
    \includegraphics[width=13cm]{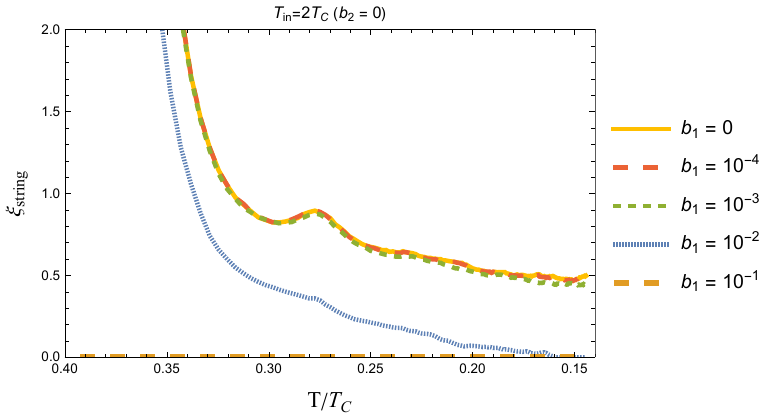}
    \caption{$N=1$ : Time evolution of $\xi_{\rm string}$ for $b_{1}=0$ (yellow solid curve), $b_{1}=10^{-4}$ (red dashed curve), $b_{1}=10^{-3}$ (green dashed curve), $b_{1}=10^{-2}$ (blue dashed curve) and $b_{1}=10^{-1}$ (orange dashed curve).}
    \label{fig:result_n1_T2}
\end{figure}

Let us now discuss cosmic string formation by numerically following the evolution of the scaling parameter, $\xi_{\rm string}$.
Figure \ref{fig:result_n1_T2} shows the evolution of the scaling parameter for some values of $b_1$ with initial temperature $T_{\rm in} = 2T_C$.
In this analysis, we generate 10 random seeds for all values of $b_1$.
It is clear that the evolution with $b_1=0$, $b_{1} = 10^{-4}$ and $b_1=10^{-3}$ is almost the same, with no significant difference within our simulation setup.
The evolution of the scaling parameter with $b_1 \gtrsim 10^{-2}$ significantly differs from that of $b_1 \lesssim 10^{-2}$.
$\xi_{\rm string}$ is suppressed for higher values of $b_1$.
For $b_1 = 10^{-1}$, $\xi_{\rm string}$ is almost zero in the temperature range $0.4>T/T_C>0.15$, and hence, there is no cosmic string formation.
For $b_1 = 10^{-2}$, there is a non-vanishing energy density of cosmic strings at high temperature $T>0.3T_C$, but it monotonically decreases in time.
Eventually, $\xi_{\rm string}$ becomes zero around $T=0.15T_C$, while it does not vanish for $b_1\leq 10^{-3}$.
At least, we confirm that $\xi_{\rm string}$ for $b_1 = 10^{-2}$ is two order magnitude smaller than that for $b_1\leq 10^{-3}$ at $T=0.15T_C$.
Therefore we conclude that cosmic strings may not be formed, or they soon disappear even right after the phase transition for $b_1 \geq 10^{-2}$.
This result is consistent with our analytic estimate Eq.~\eqref{eq:critical value}.
For $b_1\leq 10^{-3}$, cosmic strings are formed, but they may eventually decay since cosmic strings cannot sustain against the domain wall tension.
We discuss $b_1$ dependence on the time evolution of $\xi_{\rm string}$ with a larger box size $N_l= 256^3$ in App.~\ref{app:simluation with a larger box size}.

\begin{table}[]
    \centering
    \begin{tabular}{|c|c|c|c|c|}
        $b_{1}$&$10^{-1}$&$10^{-2}$&$10^{-3}$&$10^{-4}$\\\hline
        $\sqrt{\langle \delta \phi_1^2(T_C)\rangle_\beta }/|\langle\phi_{1}(T_C)\rangle|$ & 
        $0.841$&$ 1.004$&$ 1.021$&$ 1.023$
    \end{tabular}
    \caption{A ratio of the root-mean-square of the scalar field $\phi_1$ to the position of the global minimum of the thermal effective potential evaluated at $T=T_C$ are shown for some values of the explicit symmetry breaking parameter  $b_1$.}
    \label{tab:my_label}
\end{table}

In order to understand the above feature of cosmic strings formation  for $b_1 \neq 0$, we numerically investigate the corresponding semi-analytic criteria given by Eq.~\eqref{eq:string formation criterion} in the simulation.
The table \ref{tab:my_label} shows the ratio of the root-mean-square of the fluctuations of $\phi_1$ to the position of the minimum of the thermal effective potential given by Eq.~\eqref{eq:potential}.
Here all configurations are generated at the temperature $T_{\rm in}=2T_C$.
We confirm that when we change the initial temperature dependence as $T_{\rm in}=2T_C$ and $T_{\rm in} = 4T_C$, the ratio $\langle \delta\phi^2_1(T_C)\rangle_\beta /|\phi_1(T_C)|$ is changed only within $5\,\%$.
From the table \ref{tab:my_label}, we find that the rough criterion \eqref{eq:string formation criterion} works well, suggesting that cosmic strings are formed for $b_1 \lesssim 10^{-2}$ while they are not for $b_1 \gtrsim 10^{-2}$.

\begin{figure}
    \centering
    \includegraphics[width=13cm]{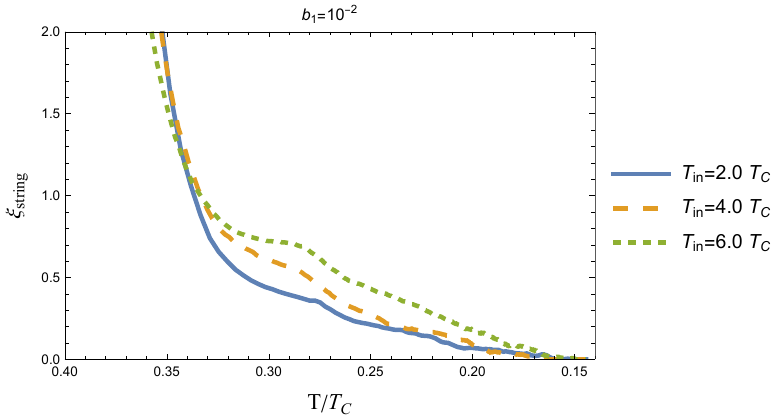}
    \caption{$N=1$ : Time evolution of $\xi_{\rm string}$ for $T_{\rm in}=2.0 \, T_{C}$ (blue solid curve), $T_{\rm in}=4.0T_{C}$ (yellow dashed curve) and $T_{\rm in}=6.0 \, T_{C}$ (green dashed curve).}
    \label{fig:result_n1_b2}
\end{figure}

One needs to discuss the effect of the initial temperature dependence on $\xi_{\rm string}$.
Figure~\ref{fig:result_n1_b2} shows the time evolution of the scaling parameter with $b_1 =10^{-2}$ for three different initial temperatures.
It is clear that $\xi_{\rm string}$ quickly converges to zero for any initial temperature although configurations generated at higher temperatures gives a higher value of $\xi_{\rm string}$ around $T>0.2\,T_C$.
Hence our conclusion is not altered by the change of the initial temperature.
This result is consistent with our earlier discussion, Eq.~\eqref{eq:initial temperature dependence}.

\subsection{The  $N=2$ case}\label{subsec:result n=2}

Here we discuss defects formation for $b_1 =0$ and $b_2\neq 0$, where there is an exact discrete symmetry $\mathbb{Z}_2$.
We prepare initial field configurations at high temperatures $T>T_{C_2}$, so that this $\mathbb{Z}_2$ symmetry is restored.
As we discussed in Sec.~\ref{subsec:setup-n2}, due to the presence of an exact $Z_2$ symmetry, there can be a composite solitonic configuration, where a single cosmic string is attached to two domain walls in the low temperature region.
The dynamics of the string-wall network is dramatically different from the $b_1\neq 0$ and $b_2=0$ case.

\begin{figure}
    \centering
    \includegraphics[width=13cm]{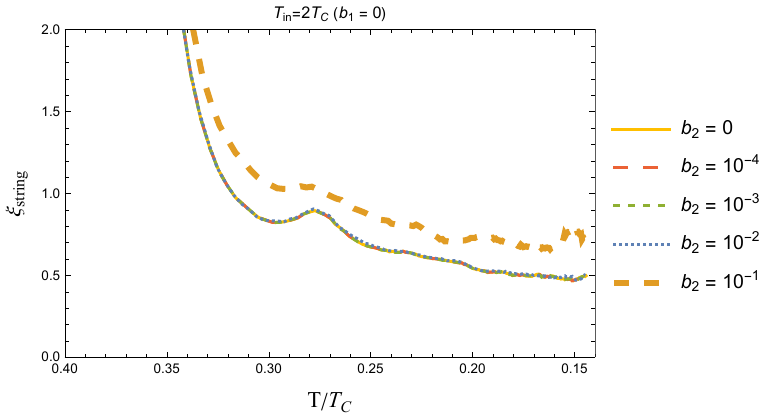}
    \caption{$N=2$ : Time evolution of $\xi_{\rm string}$ for $b_{2}=0$ (yellow solid curve), $b_{2}=10^{-4}$ (red dashed curve), $b_{2}=10^{-3}$ (green dashed curve), $b_{2}=10^{-2}$ (blue dashed curve) and $b_{2}=10^{-1}$ (orange dashed curve).}
    \label{fig:result-n2-T2}
\end{figure}

Time evolution of the scaling parameter of cosmic string is shown in Fig.~\ref{fig:result-n2-T2}.
The cosmic string energy density never vanishes at least in the region $T\geq 0.15T_C$ (and $b_2\leq 10^{-1}$).
This feature is in sharp contrast with the $N=1$ case, in which the string energy density vanishes for relatively larger breaking parameter $b_1 \gtrsim 10^{-2}$.
As we discussed in Sec.~\ref{sec:setup}, two domain walls are randomly distributed, and hence, smoothness of the field configuration unavoidably generates the cosmic string configuration as long as $b_2\leq 1$.

Let us define $T_{\rm DW}$ to be the temperature at which the domain wall tension becomes dominant compared to the string tension for a string loop whose radius is an order of the Hubble scale.
This temperature can be evaluated by comparing forces acting on the string from the domain wall tension and the cosmic string tension~\cite{Vilenkin:1982ks}, and turns out to be
$T_{\rm DW}\sim(\lambda b_{2}\zeta^{2})^{1/4}T_{C}\sim 0.2\,(b_{2}/10^{-3})^{1/4}T_{C}$.
According to this expression, a larger value of $b_2$ implies that the string-wall system is effectively governed by the wall dynamics at a higher temperature.
An interesting consequence of a large amount of $b_2 =10^{-1}$ is that $\xi_{\rm string}$ with $b_2 =10^{-1}$ is slightly larger than that with $b_2 = 0$. This may be mainly because the potential hills at $\theta=0,\pi$ prevent configurations from decaying into one of the vacua such that a larger size of $b_{2}\geq 10^{-1}$ enhances the string formation.

\begin{figure}
    \centering
    \includegraphics[width=13cm]{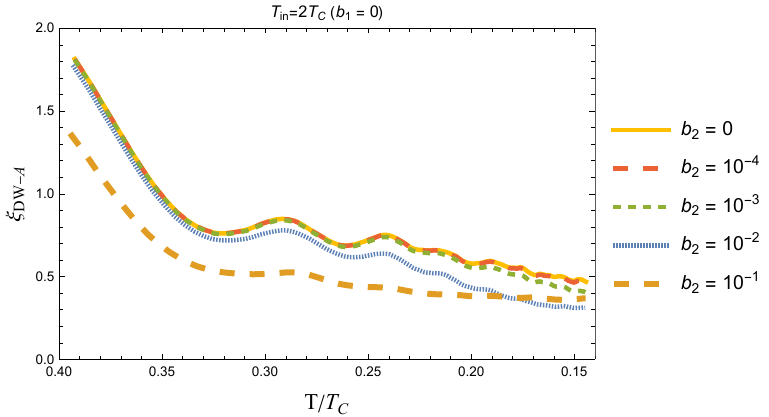}
    \caption{$N=2$ : Time evolution of $\xi_{\rm DW-A}$ for $b_{2}=0$ (yellow solid curve), $b_{2}=10^{-4}$ (red dashed curve), $b_{2}=10^{-3}$ (green dashed curve), $b_{2}=10^{-2}$ (blue dashed curve) and $b_{2}=10^{-1}$ (orange dashed curve).}
    \label{fig:result-n2-T2-DWA}
\end{figure}
\begin{figure}
    \centering
    \includegraphics[width=13cm]{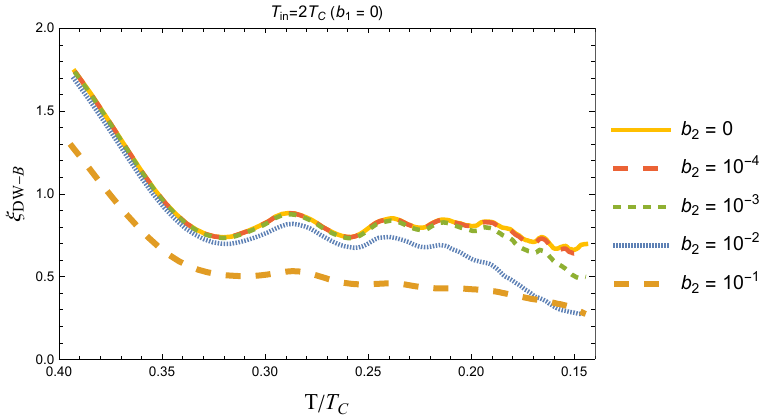}
    \caption{$N=2$ : Time evolution of $\xi_{\rm DW-B}$ for $b_{2}=0$ (yellow solid curve), $b_{2}=10^{-4}$ (red dashed curve), $b_{2}=10^{-3}$ (green dashed curve), $b_{2}=10^{-2}$ (blue dashed curve) and $b_{2}=10^{-1}$ (orange dashed curve).}
    \label{fig:result-n2-T2-DWB}
\end{figure}

We also show the time evolution of two domain wall scaling parameters in Fig.~\ref{fig:result-n2-T2-DWA}.
It is clear from the figures that both domain wall energy densities do not vanish within the simulation time available in our numerical setup, which indicates that the both domain walls are stable.
For a smaller size of $b_2<10^{-3}$, the dynamics of the string-wall network is effectively governed by the string tension rather than the domain wall tension for $T\geq 0.2T_C$.
Consequently, it is easy to overcome the potential barrier of the domain wall,
which leads to the oscillation pattern for both domain wall scaling parameters as can be seen from the figures.
For a larger size of $b_2 \geq 10^{-2}$, the domain wall tension becomes effective compared to the case with $b_2\leq 10^{-3}$, and hence, amplitudes of oscillations of domain wall scaling parameters are suppressed. These features can be understood as follows.
The scalar field configurations are well biased by the zero mode oscillation, which is suppressed for a larger size of $b_{2}\geq 10^{-2}$ because the potential hills at $\theta=0,\pi$ get higher and require more energy to overcome.


\section{Conclusions and discussion}\label{sec:discussion}

We have explored formation of cosmic strings associated with a spontaneously broken {\it approximate} global $U(1)$ symmetry by performing a lattice simulation in a radiation-dominated Universe and taking effects of thermal fluctuations into account. Let us emphasize  that this situation is different from the one when a global $U(1)$ symmetry is spontaneously broken first and then explicit symmetry breaking takes place later. This kind of situation has been often discussed in the context of axion cosmology. In our case, an explicit symmetry breaking has already occurred before spontaneous breaking of the {\it approximate} global symmetry.
As concrete examples of such a situation, we consider explicit breaking operators that break an original global $U(1)$ symmetry down to its discrete subgroup $Z_N$ for $N=1$ and $N=2$, respectively.
Although the topology of the ground state is trivial and hence there is no stable topological soliton in a strict sense in the case $N=1$, there can be a formation of composite topological soliton configuration if the effect of explicit breaking is small enough.
In particular, we find that cosmic strings (attached to a single domain wall) can be produced at the critical temperature defined in the absence of the explicit breaking parameters when the amount of the ratio of the spontaneous symmetry breaking to that of the explicit breaking is smaller than $\mathcal{O}(10^{-2})$ for $N=1$.
A salient feature of the case $N=2$ is that there exists an exact global $Z_2$ symmetry which is spontaneously broken at a certain temperature.
By studying the form of the thermal effective potential, it turns out that there exists composite solitons, where a single string attached to two domain walls for $N=2$.
Contrary to the $N=1$ case, this composite soliton formation takes place as long as the amount of the explicit breaking is smaller than that of the spontaneous breaking for $N=2$.
When the temperature of the Universe is sufficiently high enough to restore this $Z_2$ symmetry, domain walls are inevitably produced.
One of the important consequence of this work is that there may be no pNGB (majoron) production from cosmic strings when the ratio of the explicit symmetry breaking to the spontaneous symmetry breaking is greater than $10^{-2}$.
In order to precisely estimate the energy density of the cosmic string after their formation, one requires a much larger box to follow evolution of string network as well as the analysis beyond mean field approximation around the critical temperature, which are the beyond scope of the present paper.


\section*{Acknowledgments}
We would like to thank the participants of the workshop "IBS CTPU-CGA 2023 Workshop on Topological Defects" held at the IBS in Daejeon for useful comments and discussions. In particular, we would like to thank Muneto Nitta and Arttu Rajantie for valuable discussions.
K.F. is supported by JSPS Grant-in-Aid for Research
Fellows Grant No.22J00345. M.S. is partially supported by the Science and Technology Facilities Council (STFC grant ST/X000753/1).
M.U. is supported by JSPS Grant-in-Aid for Research Fellows Grant No.24KJ1118 and by IBS under the project code, IBS-R018-D3. M.Y. is supported by IBS under the project code, IBS-R018-D3, and by JSPS Grant-in-Aid for Scientific Research Number JP21H01080.


\appendix

\section{Scalar field oscillation on an expanding background}\label{app:scalar field oscillation}

Let us discuss time evolution of the amplitude of a homogeneous real scalar field $\phi(t)$ minimally coupled to gravity with temperature dependent potential $V(\phi,T)$ on an expanding background whose metric is given by Eq.~\eqref{eq:metric}.

The equation of motion of the real scalar field is 
\begin{align}
    \ddot{\phi}+3H\dot{\phi}+\frac{\partial V}{\partial \phi}=0.\label{eq:eom of a scalar field}
\end{align}
Let us specify the form of the potential $V(\phi(t),T(t))= c_1T^2(t)\phi^2(t)$ with a positive constant $c_1$; the temperature depends on the cosmic time.
This form is relevant in our situation, but an extension to a more complicated expression is straightforward.
Suppose that the real scalar field is initially displaced from its potential minimum and starts to oscillate.
In the following discussion, we assume that the frequency of the oscillating scalar field is much larger than the Hubble parameter, $\sqrt{c_1}T(t)\gg H(t)$.

Since we are interested in the change of the time scale which is much longer than the single period of the oscillating scalar field represented by $P$, we  define the average of the function $f(t)$ over a single period as
\begin{align}
    \langle f(t)\rangle _{P} \equiv \frac{1}{P}\int _{t}^{t+P}f(t')\,\mathrm{d}t'.
\end{align}
Then we obtain
\begin{align}
        \langle K\rangle =\left\langle\frac{1}{2}\dot{\phi}^{2}\right\rangle _{P}
        =-\frac{1}{2P}\int_{0}^{P}\phi\ddot{\phi}dt
        =\langle V \rangle,
\end{align}
In the first line, we perform the integration by parts with vanishing surface term.
In the last equality, we use the equation of motion given in Eq~\eqref{eq:eom of a scalar field} by assuming $H(t)={\rm  const}$ and neglecting the friction term during a time interval $P$.
This is a familiar expression of the virial theorem.

One can also show that the energy density of the oscillating scalar field $\rho_\phi\equiv K+V$ satisfies
\begin{align}
    \begin{split}
        \frac{\mathrm{d}}{\mathrm{d}t}\rho_{\phi} 
        &= -3H\dot{\phi}^{2} -2HV.
    \end{split}
\end{align}
Here, we use Eq.~\eqref{eq:eom of a scalar field}, $T(t)\propto a^{-1}(t)$ assuming a negligible change of the effective number of entropy degrees of freedom, and $H=1/(2t)$ which is the case for the radiation-dominated Universe.
A remarkable feature is that the second term of the right hand side of the above equation comes from the time dependence of the cosmic temperature.
By taking the average of the above equality, we find 
\begin{align}
    \begin{split}
         \frac{\mathrm{d}}{\mathrm{d}t}\left\langle \rho_{\phi}(t)\right\rangle_{P}=\left\langle \frac{\mathrm{d}}{\mathrm{d}t}\rho_{\phi}(t)\right\rangle_{P}=-4\langle \rho_{\phi}(t)\rangle_{P}.
    \end{split}
\end{align}
Thus, we obtain $\rho_\phi(t)\propto a^{-4}(t)$ corresponding to $\phi(t)\propto a^{-1}(t)$.

\section{Comparison with larger simulation}\label{app:simluation with a larger box size}
As discussed in Ref.~\cite{Yamaguchi:2002sh}, defects start to decay due to the periodic boundary condition when the volume of the simulation box becomes comparable to the Hubble volume. 
This effect is also confirmed in this study for $T\lesssim 0.15T_{C}$, as discussed in Sec.~\ref{subsec:result SSB}, requiring a larger simulation box to avoid this.
In the following, we present the results of numerical simulations with changing $N_{l}$ from $128^{3}$ to $256^{3}$ and maintaining the other parameters presented in Sec.~\ref{sec:numerical simulation}. 
Since the simulation cost increases in proportion to the computational time, we fix $T_{\rm in}=2T_{C}$  and leave the simulation for higher initial temperatures for future work. 

\begin{figure}
    \centering
    \includegraphics[width=13cm]{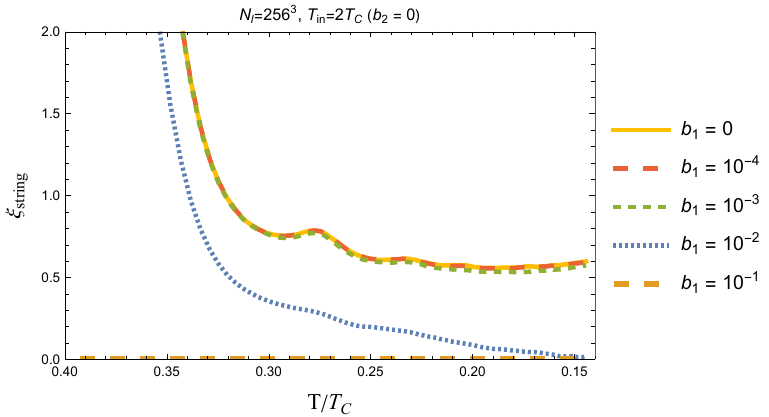}
    \caption{The time evolution of $\xi_{\rm string}$ simulated within the box $N_{l}=256^{3}$ for $b_{1}=0$ (yellow solid curve), $b_{1}=10^{-4}$ (red dashed curve), $b_{1}=10^{-3}$ (green dashed curve), $b_{1}=10^{-2}$ (blue dashed curve) and $b_{1}=10^{-1}$ (orange dashed curve).}
    \label{fig:result_n1_T2_256}
\end{figure}

Figure \ref{fig:result_n1_T2_256} shows the time evolution of $\xi_{\rm string}$ for $N=1,\, T_{\rm in}=2T_{C},\,N_{l}=256^{3}$. In this plot, $\xi_{\rm string}$ is almost constant for $T\leq 0.2T_{C}, \, b_{1}\leq 10^{-3}$ due to the absence of the finite volume effects, which is clearly different from Figure \ref{fig:result_n1_T2} where $\xi_{\rm string}$ keeps decreasing even for $b_{1}=0$. However, the overall behavior is consistent. For $b_{1}\leq 10^{-3}$, the scaling parameters are almost same. For $b_{1}=10^{-2}$, $\xi_{\rm string}$ does not vanish at $T>T_{C}$, but it monotonically decreases in time and becomes zero around $T=0.15 T_{C}$. $\xi_{\rm string}$ is almost zero in the temperature range $0.4 > T/T_{C} > 0.15$. Therefore, we conclude that the argument in Sec.~\ref{subsec:result n=1} is universal without suffering from the finite volume effects. 

\begin{figure}
    \centering
    \includegraphics[width=13cm]{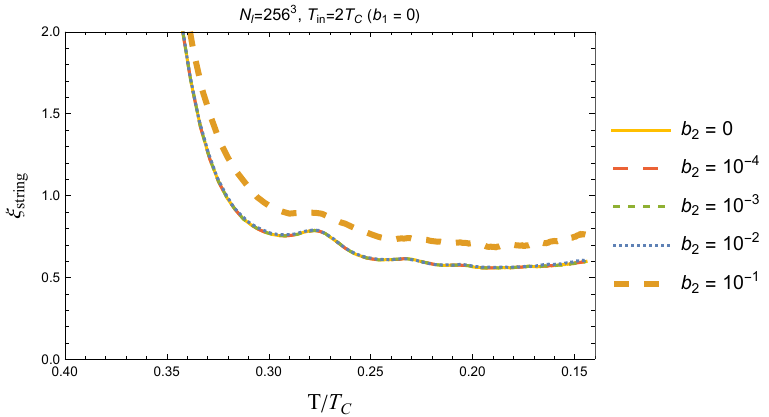}
    \caption{Time evolution of $\xi_{\rm string}$ simulated within the box $N_{l}=256^{3}$ for $b_{2}=0$ (yellow solid curve), $b_{2}=10^{-4}$ (red dashed curve), $b_{2}=10^{-3}$ (green dashed curve), $b_{2}=10^{-2}$ (blue dashed curve) and $b_{2}=10^{-1}$ (orange dashed curve).}
    \label{fig:result_n2_T2_256}
\end{figure}

\begin{figure}
    \centering
    \includegraphics[width=13cm]{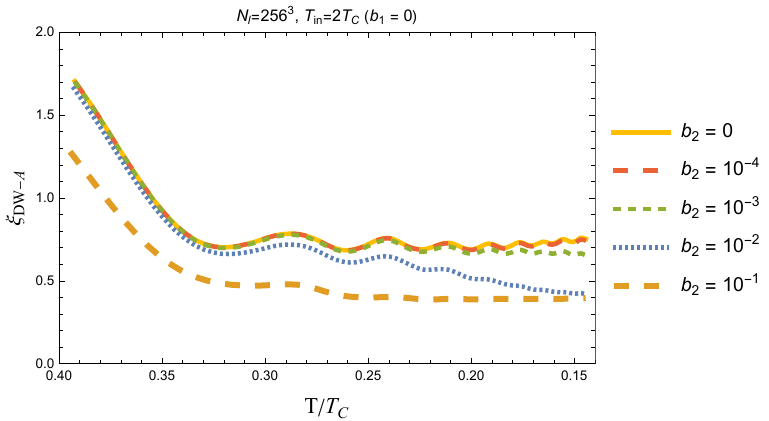}
    \caption{The time evolution of $\xi_{\rm DW-A}$ simulated within the box $N_{l}=256^{3}$ for $b_{2}=0$ (yellow solid curve), $b_{2}=10^{-4}$ (red dashed curve), $b_{2}=10^{-3}$ (green dashed curve), $b_{2}=10^{-2}$ (blue dashed curve) and $b_{2}=10^{-1}$ (orange dashed curve).}
    \label{fig:result_n2_T2-DWA_256}
\end{figure}

\begin{figure}
    \centering
    \includegraphics[width=13cm]{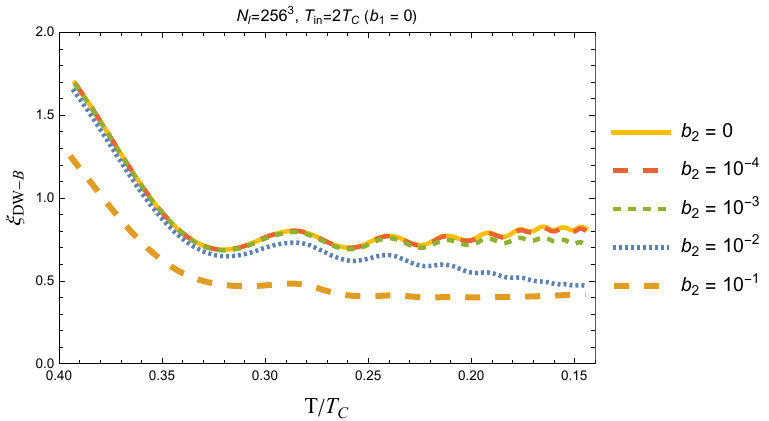}
    \caption{Time evolution of $\xi_{\rm DW-B}$ simulated within the box $N_{l}=256^{3}$ for $b_{2}=0$ (yellow solid curve), $b_{2}=10^{-4}$ (red dashed curve), $b_{2}=10^{-3}$ (green dashed curve), $b_{2}=10^{-2}$ (blue dashed curve) and $b_{2}=10^{-1}$ (orange dashed curve).}
    \label{fig:result_n2_T2-DWB_256}
\end{figure}

We also confirm that argument on formation of string-wall system  for $N=2$ obtained in Sec.~\ref{subsec:result n=2} is universal without suffering from the finite volume effects.
Figures \ref{fig:result_n2_T2_256}, \ref{fig:result_n2_T2-DWA_256}, and \ref{fig:result_n2_T2-DWB_256} show the time evolution of $\xi_{\rm string}$, $\xi_{\rm DW-A}$ and $\xi_{\rm DW-B}$ for $N=2,\, T_{\rm in}=2T_{C},\,N_{l}=256^{3}$, respectively. The overall behavior is again consistent with the results obtained in Sec.~\ref{subsec:result n=2}. More precisely, $\xi_{\rm string}$ is almost the same for $b_{2}\leq 10^{-2}$ and slightly amplified for $b_{2}=10^{-1}$. $\xi_{\rm DW-A, B}$ is almost the same for $b_{2}\leq 10^{-3}$ where the dynamics of the string-wall network is effectively governed by the string tension. For a larger size of $b_{2}\geq 10^{-2}$ where the domain wall tension becomes effective, the amplitude of the oscillation of $\xi_{\rm DW-A, B}$ is suppressed.


\bibliography{reference} 
\bibliographystyle{JHEP}

\end{document}